\documentclass{aa}
\usepackage{natbib}
\usepackage{graphicx}
\bibpunct{(}{)}{;}{a}{}{,} 

\begin{document}
\title{X-ray emission from classical T Tauri stars: Accretion shocks and coronae? \thanks{Based on observations obtained with XMM-Newton, an ESA science mission with instruments and contributions directly funded by ESA Member States and NASA.}}

\author{H.M. G\"unther \and J.H.M.M. Schmitt \and J. Robrade \and C. Liefke}  
\offprints{H. M. G\"unther,\\ \email{moritz.guenther@hs.uni-hamburg.de}}
\institute{Hamburger Sternwarte, Universit\"at Hamburg, Gojenbergsweg 112, 21029 Hamburg, Germany}
\date{Received 23 May 2006 / accepted 21 February 2007}
\abstract{Classical T Tauri stars (CTTS) are surrounded by actively accreting disks.  According to current models material falls along the magnetic field lines from the disk with more or less free-fall velocity onto the star, where the plasma heats up and generates X-rays.}{We want to quantitatively explain the observed high energy emission and measure the infall parameters from the data. Absolute flux measurements allow to calculate the filling factor and the mass accretion rate.}{We use a numerical model of the hot accretion spot and solve the conservation equations.
}{A comparison to data from XMM-Newton and \emph{Chandra} shows that our model reproduces the main features very well. It yields for TW~Hya a filling factor of 0.3~\% and a mass accretion rate $2\times10^{-10}\ M_{\sun}\; \mathrm{ yr}^{-1}$.}{}
\titlerunning{Modelling the accretion in CTTS}
\maketitle

\section{Introduction}

T Tauri stars are young ($<10\;\mathrm{Myr}$), low mass ($M_*<3M_{\sun}$), pre-main sequence stars exhibiting strong H$\alpha$ emission.
The class of "Classical T Tauri stars" (CTTS) are 
surrounded by an accretion disk and are actively accreting material from the disk.
The disks do not reach all the way to the stellar surface, rather they are truncated in the vicinity of the corotation radius. Infrared (IR) observations 
typically yield inner radii of 0.07-0.54~AU \citep{2003ApJ...597L.149M}, consistent with the corotation radius. Disk material is ionised by energetic stellar radiation and -- somehow -- loaded onto the stellar magnetic field lines, traditionally assumed to be dipole-like \citep[but see][]{Valenti04,2006MNRAS.371..999G}.  Along the magnetic field accretion funnels or curtains develop and matter impacts onto the star at nearly free-fall velocity \citep{1984PASJ...36..105U,1991ApJ...370L..39K}. This process can remove angular momentum from the star \citep{1994ApJ...429..781S}.
Observationally the accretion can be traced in the optical in the H$\alpha$ line profile \citep{2000ApJ...535L..47M}, in the 
IR \citep{2001ApJ...551.1037B} and in the ultraviolet UV \citep{2005AJ....129.2777H}. Further support for this scenario comes from the measurement of
magnetic fields in some CTTS using the technique of spectropolarimetry and Zeeman-broadening  \citep{1999ApJ...516..900J,2003RMxAC..18...38J,2005MNRAS.358..977S}. 
The kinetic energy of the accretion stream is released in  one or several hot spots close to the stellar surface, producing the observed veiling continuum, 
and also line emission in the UV and X-rays. The emitted UV continuum radiation was previously calculated by \citet{calvetgullbring} and detailed models of the accretion 
geometry prove that stable states with two or more accretion spots on the surface can exist \citep{2004ApJ...610..920R}. The UV emission has been also used to estimate mass accretion rates in CTTS with typical values of $10^{-8}M_{\sun}$~year$^{-1}$ \citep{1999MNRAS.304L..41G,2000ApJ...539..815J}.

T Tauri stars are copious emitters of X-ray emission. Specifically, X-ray emission from quite a few  CTTS was detected with the \emph{Einstein} \citep{1981ApJ...243L..89F,1989ApJ...338..262F} 
and \emph{ROSAT} satellites \citep{1993ApJ...416..623F,1995A&A...297..391N,1998A&A...331..193G}. 
The origin of the detected X-ray emission is usually interpreted as a scaled-up version of coronal activity as observed for our Sun, and
the data and their interpretation prior to the satellites XMM-Newton and \emph{Chandra} are summarised in a review by \citet{1999ARA&A..37..363F}. 
However, recent observations of the CTTS \object{TW~Hya} \citep{2002ApJ...567..434K,twhya}, \object{BP~Tau} \citep{bptau} and
\object{V4046~Sgr} \citep{v4046} 
with the grating spectrometers onboard XMM-Newton and \emph{Chandra}
indicate very high plasma densities in the X-ray emitting regions much higher than those observed in typical coronal 
sources.  This finding suggests a different origin of at least
the soft part of the X-ray spectrum in CTTS.  Simple estimates show that X-rays can indeed be produced in the accretion spot
of a typical CTTS \citep{lamzin}. We present here a more detailed accretion shock model, 
which predicts individual emission lines and can thus be directly confronted with observations to determine model parameters
such as the maximum plasma temperature and mass accretion rate.

Unfortunately, only a few CTTS have so far been studied in detail using high-resolution 
X-ray spectroscopy with sufficient signal-to-noise ratio. It is therefore unknown at present whether the observed
low forbidden to intercombination line ratios in He-like ions as measured for the CTTS TW~Hya, BP~Tau and V4046~Sgr 
are typical for CTTS as a class.  Lower resolution studies of CTTS show
significant differences between individual stars, possibly caused by
considerable individual coronal activity, and in general it is difficult to disentangle coronal and accretion contributions \citep{Rob0507} in the
X-ray spectra of CTTS.
Other suggested sources of X-rays include dense clumps in the stellar or disk wind of CTTS, heated up by shock waves. Simulations by \citet{2002ApJ...574..232M} show sufficiently dense regions in the stellar magnetosphere.

The goal of this paper is to consider the maximally possible accretion shock
contribution, to determine the physical shock parameters, compute the emitted X-ray spectrum  and assess to what extent other
emission components are necessary to account for the observed X-ray spectra.
The detailed plan of our paper is as follows: In Sect.~\ref{obs} the observations used in this study are described briefly, 
in Sect.~\ref{model} we present our model and give the main assumption and limitations, in Sect.~\ref{res} the results of the 
simulation are shown and applied to observational data, followed by a short discussion of the main points in Sect.~\ref{dis}.

\section{Observations}
\label{obs}
\subsection{Stellar parameters
}
With a distance of only 57~pc \citep{1998MNRAS.301L..39W} TW~Hya is the closest known CTTS; it is not submerged in a dark, molecular cloud like many other CTTS. 
Photometric observations show variability between magnitude 10.9 and 11.3 in the V-band \citep{1983A&A...121..217R}. Broad H$\alpha$ profiles (FWHM $\sim 200$~km~s$^{-1}$) were observed by \citet{2000ApJ...535L..47M}, and TW~Hya apparently belongs to a group of objects with similar age, the so-called TW~Hydrae association \citep[TWA;][]{1999ApJ...512L..63W}. TW~Hydrae's mass and radius are usually quoted as $M_*$=0.7~$M_{\sun}$, $R_*$=1.0~$R_{\sun}$, and its age as 10~Myr from \citet{1999ApJ...512L..63W}. Alternative values are given by \citet{2002ApJ...580..343B}, who place TW~Hya on the HR diagram by \citet{1998A&A...337..403B} and determine stellar parameters from the optical spectrum, which fits an older (30~Myr) and smaller star ($R_*$=0.8~$R_{\sun}$). The spectral type of TW~Hya is K7$\;$V-M1$\;$V \citep{1999ApJ...512L..63W,2002ApJ...580..343B} and the system is seen nearly pole-on \citep{1997Sci...277...67K, 2000ApJ...534L.101W, 2002ApJ...571..378A}.  Moreover, TW~Hya displays variations in line profiles and veiling, which have been interpreted as signatures of accretion spot rotation \citep{2002ApJ...571..378A,2002ApJ...580..343B}. TW~Hya has been observed in the UV with \emph{IUE}, \emph{FUSE} and \emph{HST/STIS} \citep{2002ApJ...572..310H}, revealing a wealth of H$_2$ emission lines, consistent with the origin in the surface of a irradiated disk, and in X-rays with \emph{ROSAT} by \citet{2000A&A...354..621C}, \emph{Chandra}/HETGS \citep{2002ApJ...567..434K} and XMM-Newton/RGS \citep{twhya}, where the grating data indictes a significant accretion shock contribution. 

\subsection{X-ray observations}

\label{obs_data}We use high resolution spectra obtained with the \emph{Chandra} and XMM-Newton.
TW~Hya was observed for 48~ks with the \emph{Chandra} HETGS on July 18, 2000 (Chandra ObsId 5). \citet{2002ApJ...567..434K} report atypically high densities measured from the \ion{O}{vii} and \ion{Ne}{ix} triplets, and a very high neon abundance as observed for many active coronal sources in combination with a low iron abundance; the anomalously high neon abundance of TW~Hya was investigated in more detail by \citet{2005ApJ...627L.149D}. A different approach to assess the plasma density of the emitting material by means of iron line ratios was performed by \citet{Ness0510}. First-order grating spectra were extracted applying standard CIAO 3.2 tools, positive and negative orders were added up. Individual emission lines in the HEG and MEG spectra were analysed with the CORA line fitting tool \citep{2002AN....323..129N}, assuming modified Lorentzian line profiles with $\beta = 2.5$. A flare occurring in the second half of the observation was already mentioned by \citet{2002ApJ...567..434K}; for our global fitting approach we excluded the flaring period to avoid contamination due to the probable coronal origin of the flare.

Another X-ray spectrum was taken with XMM-Newton on July 9, 2001 with an exposure time of 30~ks (Obs-ID 0112880201) using the RGS as prime instrument. An analysis of this observation was presented in \citet{twhya}. We newly reduced also this dataset with the XMM-Newton Science Analysis System (SAS) software, version 6.0 and applied the standard selection criteria. The X-ray spectral analysis was carried out using XSPEC V11.3 \citep{1996ASPC..101...17A}, 
and CORA for line-fitting. Because the line widths are dominated by instrumental broadening we keep them fixed at $\Delta\lambda=0.06$~\AA{}. The RGS spectra cover a larger wavelength range than HETGS spectra and include, in addition to the He-like triplets of Ne and O, also the N triplet. Both observations show the observed helium-like triplets to be incompatible with the low-density limit and an emission measure analysis indicates the presence of plasma with a few million degrees emitting in the soft X-ray region \citep{2002ApJ...567..434K,twhya}.  

\section{The model}

\label{model}
\begin{figure}
\resizebox{\hsize}{!}{\includegraphics{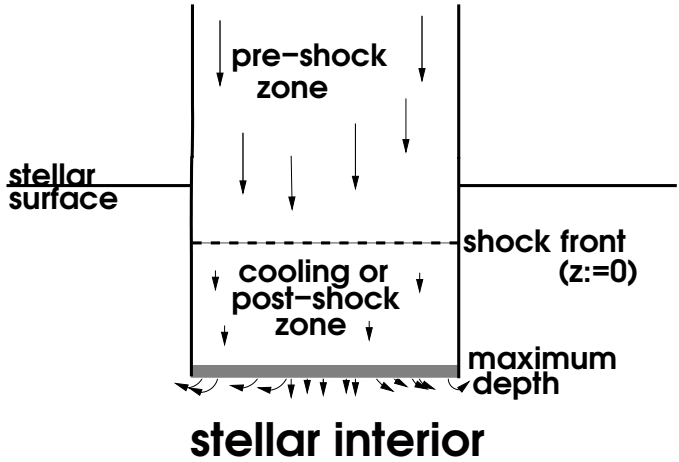}}
\caption{Sketch illustrating the structure on the accretion column. In the stellar atmosphere a thin standing shock front forms followed by a radiative cooling zone.}
\label{funnel_names}
\end{figure}

In the currently accepted accretion paradigm the material follows the magnetic field lines from the disk down to the surface of the star. Here we only model the base of the accretion column, where the infalling material hits the stellar surface, is heated up in a shock and cools down radiatively.  A sketch of the envisaged accretion scenario is shown in Fig.~\ref{funnel_names}. 
Calculations of the excess continuum produced in this region
were carried out by \citet{calvetgullbring}. \citet{lamzin} already computed the emerging soft X-ray emission from such shocks, but due to bin sizes of 50~\AA{} his results do not resolve individual lines, thus concealing much of the valuable diagnostic information. Nevertheless his models show that the hot spot produced by accretion can possibly produce not only the veiling, but also the soft X-ray emission. In our modelling we resolve all individual emission lines, allowing us to use line ratios as sensitive tracers of the density and temperature in the emitting region, and determine the elemental abundances of the emitting regions. Additionally we explicitly consider non-equilibrium ionisation states and distinguish ion and electron temperatures.\\
We assume a one-dimensional, plane parallel geometry for the accretion column. This is reasonable because the filling factor $f$, expressing the ratio between total spot size $A_{\mathrm{spot}}$ and stellar surface $A_*=4\pi R_*^2$, is quite small as will be demonstrated a posteriori. The magnetic field is assumed to be perpendicular to the stellar surface and the material flows along the magnetic field lines. 
All turbulent fluxes are neglected and our model further assumes a stationary state. We use a two-fluid approximation attributing different temperatures to the atom/ion $T_{\rm{ion}}$ and to the electron components $T_{\rm{e}}$ of the plasma. However, both components move with the same bulk velocity $v$ because they are strongly coupled by the microscopic electric fields. We ignore energy transport by heat conduction which is justified again a posteriori (see Sect.~\ref{heat_cond}) and any radiative transport (see Sect.~\ref{opticaldepth}).

\subsection{Calculation of shock front}

Once the infalling material impacts onto the stellar surface, a shock forms, defining the origin of the depth coordinate $z$ (see Fig.~\ref{funnel_names}). The shock front itself is very thin, only of the order of a few mean free paths \citep{raizerzeldovich}, thus contributing only very marginally to the total emission. Therefore it is not necessary to numerically resolve its internal structure, rather it can be treated as a mathematical discontinuity and the total change in the hydrodynamic variables over this discontinuity is given by the Rankine-Hugoniot conditions \citep[][chap.~7, \S~15]{raizerzeldovich}, which transform supersonic motion into subsonic motion in one single step. To simplify the numerical treatment we assume the direction of flow to be exactly parallel the magnetic field lines, so the Lorentz force does not influence the dynamics; we also expect the magnetic field to suppress heat conduction. Following the treatment by \citet{calvetgullbring} and \citet{lamzin} we assume the gas to expand into a vacuum behind 
the shock front. Because of larger viscous forces the strong shock formation occurs only in the ionic component, while
the electron component is first only adiabatically compressed and subsequently heated through electron-ion collisions. In order 
to calculate the state of the ionic plasma behind the shock front in terms of the pre-shock conditions only
the fluxes of the conserved quantities mass, momentum and energy are required. Marking the state in front of the shock front by the index 0, that behind the shock by index 1, the Rankine-Hugoniot conditions become
\begin{eqnarray}
\rho_0 v_0 &=& \rho_1 v_1 \label{RH1}\\
P_0+\rho_0 v_0^2 &=& P_1+\rho_1 v_1^2 \label{RH2}\\
\frac{5 P_0}{2\rho_0}+\frac{v_0^2}{2}&=&\frac{5 P_1}{2\rho_1}+\frac{v_1^2}{2} \ ,\label{RH3}
\end{eqnarray}
where $\rho$ denotes the total mass density of the gas and $P$ its pressure. 

Requiring that the electric coupling between ions and electrons leads an adiabatic compression of the electron component, implies a temperature rise
for the electronic plasma component of 
\begin{equation}
T_{e_1}=T_{e_0} \left(\frac{\rho_1}{\rho_0}\right)^{(\gamma-1)}
\end{equation}
with $\gamma=5/3$ denoting the adiabatic index. Because the time scale for heat transfer from ions to electrons is much larger than that of the ions
passing through the shock front, ions and electrons leave the shock with vastly different temperatures. Numerical evaluation of the above equations behind the accretion shock front results in electron temperatures orders of magnitude lower than the ion temperature. Furthermore,
the ions pass the shock front so fast that other degrees of freedom than kinetic are not excited and therefore kinetic and ionisation temperatures of the ions substantially differ. 

\subsection{Structure of the post-shock region}

In the following section we compute how the originally different kinetic temperatures of ions and electrons as well as the ionisation temperature
equilibrate and calculate the emitted X-ray spectrum.

\subsubsection{Momentum balance}\label{hydrodyn}

In the post-shock region
heat is transferred from the ionic to the electronic component, at the same time the gas radiates and cools down, so the energy of the gas is no longer conserved.  However, the particle number flux $j$ of ions (and atoms) 
\begin{equation}j=nv\label{j_n}\end{equation}
is conserved, where $n$ is the ion/atom number density; the electron number density is denoted by $n_{\mathrm{e}}$. The total momentum flux $j_p$ is conserved, since we ignore the momentum loss by radiation; it consists of the ion and the electron momentum as follows:
\begin{eqnarray}  
j_p&=&\mu m_{\mathrm{H}} n v^2+P_{\mathrm{ion}}+m_{\mathrm{e}} n_{\mathrm{e}} v^2+P_{\mathrm{e}} \nonumber \\
   &=&\mu m_{\mathrm{H}} n v^2+nkT_{\mathrm{ion}}+m_{\mathrm{e}} n_{\mathrm{e}} v^2+n_{\mathrm{elec}}kT_{\mathrm{e}} \nonumber\\
   &\approx&\mu m_{\mathrm{H}} n v^2+nkT_{\mathrm{ion}}+n_{\mathrm{elec}}kT_{\mathrm{e}} \ (m_{e}\ll m_{\mathrm{H}})\label{j_p}
\end{eqnarray}
with $P_{\mathrm{ion}}$ and $P_{\mathrm{e}}$ denoting the thermodynamic pressure of the ion and the electron gas respectively, $T_{\mathrm{ion}}$ and $T_{\mathrm{e}}$ their temperatures; $m_{\mathrm{H}}$ denotes the mass of a hydrogen atom, 
$m_{\mathrm{e}}$ the electron mass and $\mu$ is the dimensionless atomic weight.\\

\subsubsection{Energy balance}
\label{energybalance}

Let us next consider the energy balance in the post-shock region.  Both for the ions and the electrons
the evolution of the energy per particle is described by an ordinary differential equation (ODE).
Starting from the thermodynamic relation 
\begin{equation} \label{tsminuspdvisdu} T d\Sigma -P dV=dU \end{equation}
where $\Sigma$ denotes the entropy and $U$ the internal energy of the plasma
per heavy particle we will derive this ODE for the ionic component. 
In eqn.~\ref{tsminuspdvisdu} the quantity $T d\Sigma=dQ$ denotes the heat flux through the boundaries of the system. The system "ions" looses heat 
by collisions with the colder electrons. Heat transfer is most efficient for the lighter ions and especially protons have a much larger collision cross sections than neutral hydrogen and they are by far the most abundant species in the plasma.
To describe the heat transfer between ions and electrons we follow  \citet[Chapter VII, \S 10]{raizerzeldovich}, who give 
the heat flow $\omega_{ei}$ per unit volume per unit time as (in cgs-units)
\[
\omega_{ei}=\frac{3}{2}k n_{\mathrm{\ion{H}{ii}}} n_{\mathrm{e}} \frac{T_{\mathrm{ion}}-T_{\mathrm{e}}}{T_{\mathrm{e}}^{3/2}}\frac{\Lambda}{252} \textnormal{  (T in K)}
\]
where $k$ is Boltzmann's constant and $\Lambda$ is the Coulomb-logarithm 
\begin{equation}\Lambda\approx 9.4+1.5 \ln T_{\mathrm{e}} -0.5 \ln n_{\mathrm{e}}\ .
\label{hyd_coulomblogarithm}
\end{equation}

The number density of hydrogen ions is the number density of all heavy particles $n$ multiplied by the abundance $\xi_{\mathrm{H}}$ of hydrogen and the hydrogen ionisation fraction $x_{\mathrm{H}}^1$ :
\begin{equation} \label{omegaei}
\omega_{ei} =\frac{3}{2}k\xi_H x_H^1 n_{\mathrm{e}} \frac{T_{\mathrm{ion}}-T_{\mathrm{e}}}{T_{\mathrm{e}}^{3/2}}\frac{\Lambda}{252}\textnormal{  (T in K).}
\end{equation}
Transforming Eq.~(\ref{tsminuspdvisdu}) written per heavy particle and taking the time derivative results in
\[
\frac{dU}{dt}+P\frac{dV}{dt}=T\frac{d\Sigma}{dt}=\frac{dQ}{dt}=\omega_{ei} \ .
\]
Using the stationarity condition $\frac{d}{dt}=\frac{\partial}{\partial t}+\frac{\partial z}{\partial t}\frac{\partial}{\partial z}=v\frac{\partial}{\partial z}$ transforms this into an ODE with the dependent variable $z$, measured from the shock front inwards (see Fig.~\ref{funnel_names}); differentiation with respect to $z$ will be indicated by $'$. Thus \[v U'+PvV'=\omega_{ei} \ .\]
The internal energy $U$ is in this case the thermal energy $U=\frac{3}{2}kT$, the pressure $P$ can be rewritten using the equation of state for a perfect gas. The specific volume $V$ is the inverse of the number density $V=\frac{1}{n}$. We thus obtain
\begin{equation}
\label{energyion}
v\left(\frac{3}{2}k T_{\mathrm{ion}}\right)'+v n k T_{\mathrm{ion}} \left(\frac{1}{n}\right)'=-\omega_{ei} \ .
\end{equation}
The derivation of a corresponding equation for the electron component is similar. Since the heat loss of the ion gas is a gain for the electrons this term enters with opposite sign, and an additional loss term $Q_{col}$ appears representing energy losses by collisions which excite or ionise a heavy particle that in turn radiates.\\
It is convenient to write the electron number density as \mbox{$n_{\mathrm{e}}=x_e n$,} with $x_e$ denoting the number of electrons per heavy particle.
\begin{equation}
\label{energyelec}
v\left(\frac{3}{2}x_e k T_{\mathrm{e}}\right)'+v x_e n k T_{\mathrm{e}} \left(\frac{1}{n}\right)'=\omega_{ei}-Q_{col} x_e n,
\end{equation} 
because $\omega_{ei}$ already includes the factor $x_e n$ by definition in eqn~\ref{omegaei}.
Thus we are left with four independent variables ($n$, $v$, $T_{\mathrm{ion}}$ and $T_{\mathrm{e}}$); therefore, given $x_e$, the system of the four hydrodynamic equations (\ref{j_n}, \ref{j_p}, \ref{energyion} and \ref{energyelec}) is closed and can be solved by numerical integration. 

According to \citet[Chapter 5.3]{spitzer} particle velocities reach a Maxwellian distribution after a few mean free path lengths. Evaluating the conditions behind the shock front we find that after a few hundred meters 
such distributions are established separately for both the ions and the electrons. 
We therefore assume that both ions and electrons each have their own individual 
Maxwellian velocity distributions throughout our simulation. This allows us to define an effective kinetic
temperature for electron-atom/ion collisions. Usually collisions are treated using the ion rest frame as reference frame, 
and we fold the kinetic velocities of ion and electron and write the resulting Maxwellian distribution with the effective temperature $T_{\mathrm{eff}}=T_{\mathrm{e}}+T_{\mathrm{ion}}\frac{m_{\mathrm{e}}}{\mu m_{\mathrm{H}}}$.
This effective collision temperature is then used to calculate the radiative loss term $Q_{col}$ with the CHIANTI~4.2 code \citep{CHIANTI,CHIANTIVI}. Because the gas may be in a non-equilibrium ionisation state the tables produced by the built-in CHIANTI rad\_loss procedure, which assumes kinetic and ionisation
temperatures equilibrated,
are not valid in this case, instead a spectrum with the current state of ionisation and effective collision temperature is calculated and integrated over all contributing wavelengths to determine the instantaneous radiative losses. For fitting purposes it is further necessary to perform the simulations with different abundances because lower metallicity significantly lowers the cooling rate and therefore the extent of the post-shock cooling zone. In this approach it has to be assumed that $Q_{col}$ is constant over each time step.

\subsubsection{Microscopic physics}
The calculation of the ionisation state is completely decoupled from the hydrodynamic equations given above. 
In each time step the density and the temperatures of both plasma components are fed from the hydrodynamic into the microscopic equations below.
We assume changes in ionisation to occur only through collisions with electrons.
Ions are ionised by electron collisions (bound-free) and recombine by electron capture (free-bound). So the number density of ionisations $r_{i\rightarrow i+1}$ per unit time from state $i$ to $i+1$ is proportional to the number density of ions $n_i$ in ionisation state $i$ and the number density of electrons $n_{\mathrm{e}}$; for convenience we leave out a superscript identifying the element in question here: 
\begin{equation}
\label{mp_ionisation}
r_{i\rightarrow i+1}=R_{i\rightarrow i+1} n_{\mathrm{e}} n_i \ .
\end{equation}
Recombination is the reverse process:
\begin{equation}
\label{mp_recombination}
r_{i\rightarrow i-1}=R_{i\rightarrow i-1}n_{\mathrm{e}} n_i \ .
\end{equation}
The quantity $R_{i\rightarrow{} j}$ is the rate coefficient describing
ionisation for $j=i+1$ and recombination for $j=i-1$. For element $z$ $R_{1\rightarrow{} 0}=R_{z+1\rightarrow{} z+2}\equiv0$ because $1$ represents the neutral atom, which cannot recombine any further, and $z+1$ the completely ionised ion which cannot lose any more electrons.  The cross section $\sigma$ for each process depends not only on the ion, but also on the relative velocity of ion and electron. 
On the one hand, the number of ions in state $i$ decreases by ionisation to state $i+1$ or recombination to state $i-1$, on the other hand, it increases by ionisations from $i-1$ to $i$ and by recombination from $i+1$. For an element with atomic number $Z$ there is thus a set of $Z+1$ equations
\begin{eqnarray}
\label{mp_rateeqn}
\frac{d{n_i}}{dt}=n_{\mathrm{e}}&(R_{i-1\rightarrow{} i} \,n_{i-1}-(R_{i\rightarrow{} i+1}\,+R_{i\rightarrow{}i-1})\,n_i\nonumber\\
&+R_{i+1\rightarrow{} i}\,n_{i+1}) \ .
\end{eqnarray}

Through the electron number density $n_{\mathrm{e}}$ the equations for all elements are coupled and together with the condition of number conservation they provide a complete system of differential equations. This system can be simplified considerably under the assumption that $n_{\mathrm{e}}$ is constant during each time step $\Delta t$. Because hydrogen and helium, the main donors of electrons, are completely ionised $n_{\mathrm{e}}$ is mainly given by the hydrodynamics. This assumption leads to one independent set of equations per element. Dividing by the number density of the element in question and using that the number density $n_Z^A$ of ions in ionisation stage $Z$ for element $A$ is $n_Z^A=n \xi^A x_Z^A$ with abundance $\xi^A$ of element A and the ionisation fraction $x_Z^A$, finally leads to
\begin{eqnarray}
\label{mp_rateeqn1}
\frac{d{x_i}}{dt}=n_{\mathrm{e}} & (R_{i-1\rightarrow{} i}\, x_{i-1}-(R_{i\rightarrow{} i+1}\,+R_{i\rightarrow{}i-1})\,x_i \nonumber \\
&+R_{i+1\rightarrow i}\,x_{i+1})\ .
\end{eqnarray}

The rate coefficients $R_{i\rightarrow j}$ are taken from \citet{mazzottaetal} for dielectronic recombination, 
for the radiative recombination and the ionisation (collisional and auto-ionisation) rate we use a code from D.A. Verner, which is available in electronic form on the web\footnote{http://www.pa.uky.edu/$\sim$verner/fortran.html}. 
We calculate only the elements with Z=1-28, considering all ionisation states for each of them.
The model is implemented in IDL (Interactive data language) and the ODEs are independently integrated using 'lsode', an adaptive stepsize algorithm, which is provided in the IDL distribution.

\subsection{Verification}

In order to check our computational procedures we considered several special cases with known analytical solutions. This includes a pure hydrogen gas with different electron and ion temperatures to test the heat transfer treatment and the ionisation of hydrogen at constant temperature. 
We compare our calculations for the collisional ionisation equilibrium to \citet{mazzottaetal}, who use the same data sources as we do without the corrections from \citet{1990Ap&SS.165...27V}. Cl is the element where the largest differences occur, for all important elements all differences are marginal at best.
In addition to these physical tests we examined which spatial resolution is possible. We use an adaptive step size sampling regions with steep gradients sufficiently densely so that none of our physical variables changes by more than 5\%. In order to keep the computation time at a reasonable level, we also enforced a minimum step size of 1~m.

\subsection{Heat conduction} \label{heat_cond}

Thermal conduction tends to smooth out temperature gradients. It is not included in our simulation and we use the models' temperature gradients to estimate its importance. According to \citet[Chapter 5.5]{spitzer} the thermal heat flux $F_{\mathrm{cond}}$ is
\begin{equation} \label{res_thermal_cond}
F_{\mathrm{cond}}=\kappa_0 T^{5/2} \frac{dT}{dz} \ .\end{equation}
Here $\kappa_0 =2\times 10^{-5}{\Lambda}^{-1}\;\textnormal{erg K}^{7/2} \mathrm{s}^{-1} \mathrm{cm}^{-1}$ is the coefficient of thermal conductivity. Comparing the thermal heat flux according to this equation to the energy flux carried by the bulk motion, it never exceeds more than a few percent of the bulk motion energy transport in the main part of the cooling zone except for the lowest density cases.  Small scale chaotic magnetic fields in the plasma are possible; they would be frozen in and expected to further suppress thermal conduction. 

\subsection{Optical depth effects}
\label{opticaldepth}
Our simulation assumes all lines to be optical thin. The continuum opacity in the soft X-ray region is small, however, we need to check line opacities. The optical depth 
$\tau(\lambda)$ for a given line can be expressed as 
\begin{equation} \tau(\lambda)=\int \kappa(\lambda) dl \ ,\end{equation}
where $l$ measures distance along the photon path and $\kappa(\lambda)$ is the local absorption coefficient, which can be computed from the oscillator strength $f$ of the line in question, the number density $n_{\mathrm{low}}$ of ions in the lower state and the line profile function $\Phi(\lambda)$:
\begin{equation} \kappa(\lambda,z)= \frac{\pi e^2}{m_{\mathrm{e}} c}f n_{\mathrm{low}}(z) \Phi(\lambda,z) \ ,\end{equation}
with $e$ being the electron charge and $c$ the speed of light. We approximate the line profile to follow a Gaussian distribution law with the normalisation $\int_0^{\infty} \Phi(\lambda,z) d\lambda=1$ at all $z$, centred at $\lambda_0$ with the width $\Delta \lambda_b(z)$:
\begin{equation} 
\Phi(\lambda,z)=\frac{1}{\sqrt{\pi}}\frac{\lambda_0(z)^2}{c\Delta \lambda_b(z)}\exp\left(-\left(\frac{\lambda-\lambda_0(z)}{\Delta \lambda_b(z)}\right)^2\right) \ ;
\end{equation}
$\lambda_0(z)$ is the wavelength at line centre. Because the shocked gas is moving into the star, but decelerating, it is Doppler-shifted with the bulk velocity $v(z)$ at depth $z$ according to $\lambda_0(z)=\lambda_{\mathrm{rest}}\left(1+v(z)/c\right)$, where $\lambda_{\mathrm{rest}}$ is the rest wavelength.
The broadening $\Delta \lambda_b(z)$ is in the case of purely thermal broadening
\begin{equation}\Delta \lambda_b(z)=\frac{\lambda_0(z)}{c}\sqrt{\frac{2kT(z)}{m_{\mathrm{ion}}}} \ ,\end{equation} 
but turbulent broadening $\Delta \lambda_t(z)$ may additionally contribute. This cannot be calculated from the 1D-hydrodynamics in our approach, it can only be included in an {\it ad-hoc} fashion. On the one hand, at the
boundaries of the accretion shock region typical turbulent flows might reach velocities comparable to the bulk motion and thus
significantly broaden the observed lines, on the other hand, the magnetic fields presumed to be present should tend to suppress flows perpendicular to the field lines. For the calculation of the optical depth we chose $\Delta \lambda_t(z)=10$~km~s$^{-1}$. If the turbulent broadening is larger, the line profiles get wider and the optical depth decreases.
All lines considered in this study are excited from the ground state. Since in collisionally-dominated plasmas almost all excited ion states decay relatively fast, we assume that all ions are in their ground state. $n_{low}$ is then the product of the ionisation fraction for the line producing ion, the abundance of that element and the total ion number density.
It is important to consider that the line centre depends on depth because of the Doppler shift due to the bulk velocity. Photons emitted at line centre in deeper regions end up in the wings of the profile in higher layers. 
Since our simulation does not include radiative transfer all lines have to be checked for optical depth effects.
The exact geometry and position of the accretion shock is still unkown, but of substantial importance for estimates the line optical depth. The radiation could either escape through the boundaries of the post-shock funnel perpendicular to the direction of flow or through the shock and the pre-shock region. If the spatial extent of any single accretion funnel is small and it is located high up in the stellar atmosphere, the optical depth in the first scenario is small and the accretion contribution to the total stellar emission is large. We select this scenario in the further discussion. If, however, the shock is buried deep in the stellar atmosphere, the radiation can only escape through the shock and the thin pre-shock gas. In this case, denpending in the infall conditions, the optical depth of resonance lines can be large compared to unity.

\subsection{Limitations by 1D}

Since our model is 1D, all emitted photons can travel only up or down, they cannot leave the accretion region sideways.
To a first approximation half of the photons is emitted in either direction.  The downward emitted photons will eventually be
absorbed by the surrounding stellar atmosphere which will be heated by this radiation. The influence of the surrounding
atmosphere on the shock region is expected to be small, since the temperature and hence the energy flux from the surrounding atmosphere into the shock region is much lower. We expect the shock structure to be well represented in 1D, but the size of the hot spots could be underpredicted because, depending on the geometrical extend of each spot, less than half of the emission escapes.

\subsection{Boundary conditions and limitations of the model}

All our simulations start with a pre-shock temperature of 20\,000~K and 
and the corresponding (stationary) equilibrium ionisation state \citep{ar85}. This choice of
temperature is motivated by studies of the photoionisation in the accretion stream by the post-shock emission \citep{calvetgullbring} and analytical considerations about the electron heat flux \citep{raizerzeldovich}.  While we ignore heat conduction in the post-shock region, across the shock front the temperature gradient is of course
large.  The electrons have then mean free path lengths much larger than the shock front extent will therefore heat up the inflowing gas. We tested the influence of
different initial conditions and found that the post-shock zone depends only marginally on the chosen initial ionisation state.
We terminate our simulations  when the temperature drops below 12\,000~K. Here the opacity begins to play an important role and the energy flux from the central core of the star is no longer negligible compared to the accretion flux. Here and just behind the shock front the accuracy is affected because the step size reaches its lower limit and rapid ionisation or recombination processes are not resolved. So the model is expected to be more accurate for ions which exist at high temperatures e.g. in the formation region of \ion{O}{vii} or \ion{O}{viii}, which are precisely those observed at X-ray wavelengths.

\subsection{Model grids}

The free fall velocity onto a star with mass $M_*$ and radius $R_*$ is
\begin{equation} v_{\textnormal{free}}=\sqrt{\frac{2GM_*}{R_*}} =  617 \sqrt{\frac{M_*}{M_\odot}}\sqrt{\frac{R_\odot}{R_*}} \frac{\textnormal{km}}{\textnormal{s}}\ . \label{infallvel} \end{equation}
Typically CTTS have masses comparable to the Sun and
radii between $R_*=1.5\;R_{\sun}$ and  $R_*=4\;R_{\sun}$ \citep{2003ApJ...597L.149M}, because they have not yet finished their main sequence contraction. 
Most CTTS have inner disk radii of 10-90 solar radii \citep{2003ApJ...597L.149M}, for the specific case of TW~Hya \citet{2006ApJ...637L.133E} find an inner disk radius of $\approx12R_{\sun}$, thus the actual infall speed can almost
reach the free-fall speed.
Previous analyses indicate particle number densities of the infalling gas of about $10^{12}$~cm$^{-3}$. We therefore calculated a grid of models with infall velocities $v_0$ varying between $200$~km~s$^{-1}$ and 
$600$~km~s$^{-1}$ in steps of $25$~km~s$^{-1}$, and infall densities $n_0$ varying between $10^{10}$~cm$^{-3}$ and $10^{14}$~cm$^{-3}$ with 13 points equally spaced on a logarithmic density scale. For each model in the grid we then calculate the emissivity for selected lines in the X-ray region using the version 5.1 of CHIANTI (\citet{CHIANTI}, \citet{CHIANTIVII}). We start out with abundances from \citet{1998SSRv...85..161G} and iterate the model fits until convergence (see Sect. \ref{res_tw_abund}).

\section{Results} \label{res}

\subsection{Structure of the post-shock region}

\begin{figure}
\resizebox{\hsize}{!}{\includegraphics{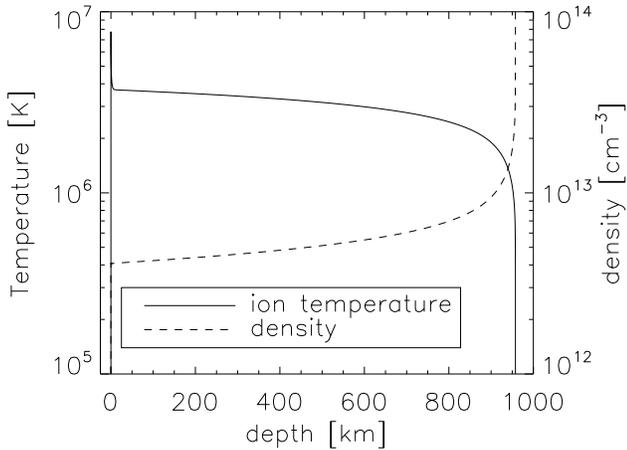}}
\caption{Temperature and density profiles for a shock calculated with $n_0=10^{12}$~cm$^{-3}$ and $v_0=525$~km~s$^{-1}$, chemical abundances as found in TW Hya (Sect. \protect{\ref{res_tw_abund}}).}
\label{res_gen_tn}
\end{figure}

\begin{figure}
\resizebox{\hsize}{!}{\includegraphics{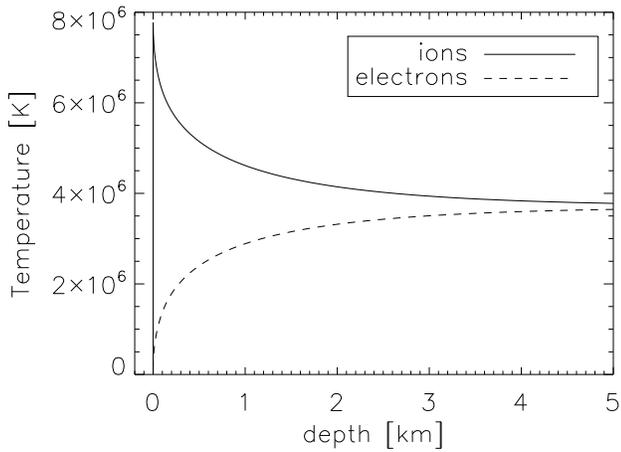}}
\caption{Ion and electron temperatures in the upper post-shock region for a shock calculated with the same starting conditions as in Fig.~\protect{\ref{res_gen_tn}}.}
\label{res_gen_t_ausgl}
\end{figure}

\begin{figure}
\resizebox{\hsize}{!}{\includegraphics{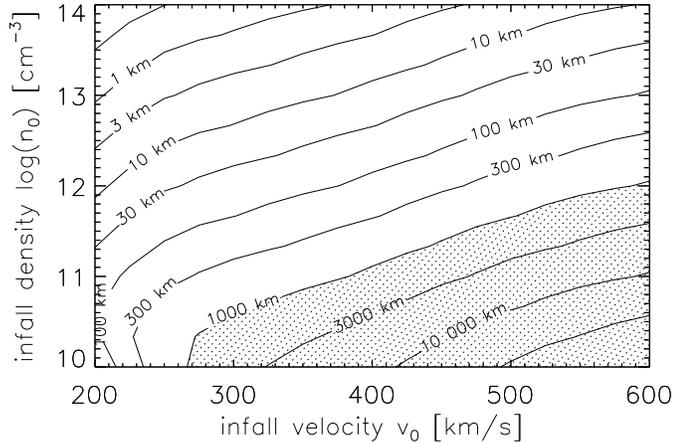}}
\caption{The length of the cooling zone dependent on the logarithm of the infall density log$n_0$ and the infall velocity $v_0$ (abundances from \citet{1998SSRv...85..161G}): The length of the cooling region is labelled. The dotted area marks regions where the deeper parts of the simulated shock are below the cut-off.}
\label{res_maxdepth}
\end{figure}

Fig.~\ref{res_gen_tn} shows typical model temperature and density profiles. In the (infinitesimally thin) shock, defined at depth 0~km, the ion temperature suddenly rises and cools down directly behind the shock front because the ion gas transfers energy to the electrons. After a few kilometers both ions and electrons have almost identical temperatures and henceforth there is essentially only radiative heat loss. This region we refer to as  the post-shock cooling zone, where most of the X-ray emission originates (see Fig.~\ref{funnel_names}).
During radiative cooling the density rises and because of momentum conservation the gas slows down at the same time (Eq.~\ref{j_p}). 
As more and more energy is lost from the system, the density and the energy loss rate increase and the plasma cools down very rapidly
in the end. In the example shown in Fig.~\ref{res_gen_tn}, the shock reaches a depth of about 950~km, which is much smaller than a stellar radius, thus justifying our simplifying assumption of a planar geometry {\it a posteriori]}.\\
The region where the electron and ion temperature substantially differ from each other is much smaller than this maximum depth, thus a two-fluid treatment is not strictly necessary for most parts of the shock. In Fig.~\ref{res_gen_t_ausgl} both temperatures are plotted in comparison. At the shock front the electrons stay relatively cool because they are only compressed adiabatically. Behind the
shock front the energy flows from the ions to the electrons, and already at a depth of ~5~km ions and electrons have almost
identical temperatures. 
\begin{figure}
\resizebox{\hsize}{!}{\includegraphics{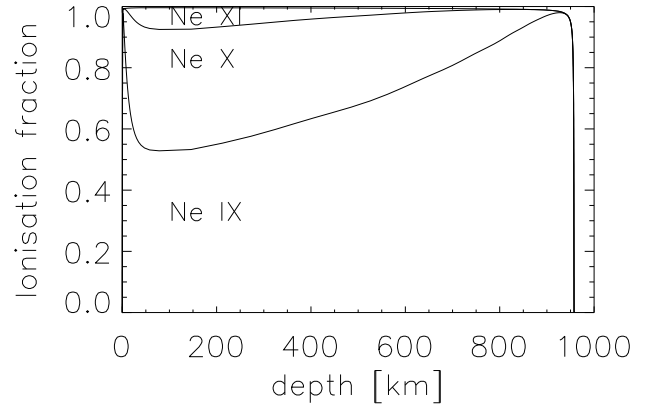}}
\caption{Ionisation state of neon for a shock calculated with the same starting conditions as in Fig.~\protect{\ref{res_gen_tn}}. See text for a detailed description.} \label{neemission}
\end{figure}
In Fig.~\ref{neemission} we show the depth dependence of the ionisation state of neon, which produces strong lines observed by {\it Chandra} and XMM-Newton. On passing through the shock front neon is nearly instantaneously ionised up to \ion{Ne}{ix}, in the following $\approx 50$~km the mean ionisation rises until the equilibrium is reached. The plasma then contains a few percent \ion{Ne}{xi} nuclei, about 40\% of the is in the form of \ion{Ne}{x}, the rest comes as \ion{Ne}{ix}. Further away from the shock the ions recombine because of the general cooling of the plasma, following the local equilibrium closely, so the fraction of \ion{Ne}{vii} rises. At the maximum depth the ions quickly recombine to lower ionisation states.\\
The maximal temperatures are proportional to $v_0^2$ as can be seen analytically from the equation of state for a perfect gas: 
\begin{equation}T_1\sim \frac{P_1}{n_1} \sim \frac{n_0 v_0^2}{n_1} \sim v_1 u_0 \sim v_0^2 \label{res_t_v} \end{equation}
These estimates are obtained using Eq.~\ref{RH1}, \ref{RH2}, \ref{RH3} and neglecting the initial pressure.\\
A higher infall velocity leads to a deeper post-shock cooling zone since the material reaches higher temperatures and consequently needs longer to cool down, so it flows for longer times and penetrates into deeper regions. 
Secondly, lower infall densities result in shocks with a larger spatial extent. Since the energy losses roughly scale with the square of the density, a lower density will increase the cooling time of the gas to cool down to photospheric temperatures. Fig.~\ref{res_maxdepth} shows that cooling zone lengths between 1~km and 10000~km can be reached depending on the chosen model parameters.
Our model assumes a free plasma flow during cooling without any direct influence from the surrounding stellar atmosphere. As long as the infalling gas has a sufficiently large ram pressure, the model assumption should be applicable, since the surrounding atmosphere is pushed away and mixes with the accreted material only after cooling. For thin gases the ram pressure is lower and the gas needs more time for cooling down. 
Therefore in the deeper and denser layers the approximation of a freely flowing gas is no longer valid and the whole set of model assumptions breaks down. The depth, where this happens, depends on the stellar parameters. 
The shock front forms where its ram pressure 
\begin{equation} \label{p_ram} p_\mathrm{ram}=\rho v_0^2 \ .\end{equation} 
roughly equals the gas pressure of the stellar atmosphere.
The pressure of the stellar atmosphere rises exponentially with depth, so, independent of the starting point, only shocks with small cooling length can be described by the hydrodynamic modelling used here. We place a cut-off at $z=1000$~km, where the pressure of the surrounding atmosphere will be larger by about an order of magnitude already.

\subsection{Optical depth}
We find that in all reasonable cases the total column density over the whole simulated region is small, in the best-fit solution if turns out to $N_H=10^{21}$~cm$^{-2}$, so the continuum opacity is small as assumed. Because emission originates at all depths the mean absorption column is even less. The line opacity along the direction of flow reaches values considerably above unity, so if the radiation passes through the whole post-shock region and does not escape through the boundaries the emssion in the resonance lines is considerably reduced (e.g. in Fig.~\ref{netriplet}). However, as we show in the following, the observed line ratios can be consistently explained in the accretion shock model, so the assumption of a geometry allowing most photons to escape, seems to be realistic.

\subsection{Application to TW~Hya}

\label{res_twhya}  In order to model the actual X-ray data available for TW~Hya we applied a two-step process: First we used only the line fluxes of selected strong lines detected in the X-ray spectra and determined the best-fit shock model in an attempt to assess the maximally possible shock contribution.  Then, in a second step, we performed a global, simultaneous fit to all available 
- lower resolution - data allowing for possible additional coronal contributions.

\subsubsection{Fit to line fluxes}

\label{res_tw_nv}
\begin{figure}
\resizebox{\hsize}{!}{\includegraphics{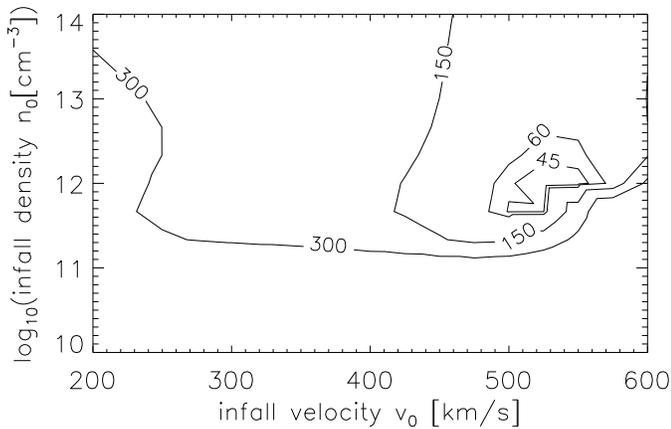}}
\caption{\label{res_fit_twhya}Contours are labelled with unreduced $\chi^2$ values for TW~Hya (7 dof), data from XMM-Newton.}
\end{figure}

Ratios between emission lines of the same element allow a determination of the best model parameters independent of the elemental abundances.  Specifically, the XMM-Newton data contain the helium-like triplets of N,O and Ne, which
strongly emit at plasma temperatures of a few MK (see Table~\ref{res_tab_chi_lines}). 
For these three elements the corresponding Ly$\alpha$ lines are also measured, while we do not use any of the Ly$\beta$ lines because they provide relatively little additional temperature sensitivity and are substantially weaker than the Ly$\alpha$ lines.  For each model we therefore compute three line ratios for each of the elements N, O and Ne, i.e., the so-called R- and G-ratios defined from
the He-like triplets \citep{1969MNRAS.145..241G,2001A&A...376.1113P} through
$R = f/i$ and $G = (f+i)/r$ respectively as well as the ratio of the 
Ly$\alpha$ to the He-like r-lines.  
These nine line ratios are compared to the
data via the $\chi^2-$ statistics; the resulting contour plot of $\chi^2$
as a function of the model parameters $n_0$ and $v_0$ is shown in 
Fig.~\ref{res_fit_twhya}.  
Correcting for the absorption does not alter the results since $N_H$ is small \citep[$N_H=3.5\cdot10^{20}\textnormal{ cm}^{-2}$;][]{Rob0507}.
The best model is found for the parameters 
$v_0=525$~km~s$^{-1}$ and $n_0=10^{12}$~cm$^{-3}$ with an unreduced $\chi^2=31.9$ (7 degrees of freedom). 
One has to be careful here in interpreting the absolute value of the $\chi^2$ because it is derived from very few highly significant numbers with non-Gaussian errors. 
 
\begin{figure}
\resizebox{\hsize}{!}{\includegraphics{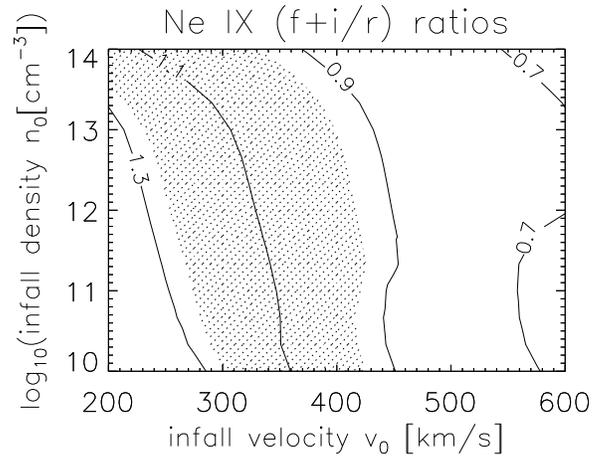}}
\caption{\label{negratio}Contour are labelled with calculates Ne G-ratios. The observed value for TW~Hya is shaded in grey (1$\sigma$-error).}
\end{figure}

The strong neon lines confine the fit most effectively because their values have the smallest statistical uncertainties. The 
density is mainly restricted by R-ratios, the velocity by the Ly$\alpha$/r-ratios, which are temperature-sensitive.
The Ne G-ratio deviates from the best fit parameters significantly (see Fig.~\ref{negratio}). 
Its observed value is $1.1\pm0.13$, which points to infall velocities between $300$~km~s$^{-1}$ and $400$~km~s$^{-1}$; fitting only the remaining eight ratios we obtain the same best fit model as before. The fit does not depend on the chosen background radiation temperature in the range 6000~K to 10000~K. 

\begin{table*}
\caption{\label{res_tab_chi_lines}The line ratios used in the fitting process for both observations are shown with their errors. The two rightmost columns show results for the best-fit scenario from the simulation with $n_0 = 10^{12} \mathrm{ cm}^{-3}$ for both models.}
\begin{center}
\begin{tabular}{lcccc}
\hline \hline
Line ratio & XMM/RGS  & Chandra/HETGS & $v_0=525$~km~s$^{-1}$ & $v_0=575$~km~s$^{-1}$ \\
\hline
                  N R-ratio          &$ 0.33 \pm 0.24 $ & n. a. &0.00 &0.00\\
                   N G-ratio         &$ 0.88 \pm 0.31 $ & n. a. &0.77 &0.75\\
N Ly$\alpha$/\ion{N}{vi} r            &$ 4.04 \pm 1.00 $ & n. a. &2.40 &2.82\\    
                  O R-ratio          &$ 0.06 \pm 0.05 $ &$ 0.04 \pm 0.06$ &0.02 &0.02 \\
                  O G-ratio          &$ 0.51 \pm 0.07 $ &$ 0.82 \pm 0.22$ &0.73 &0.71\\
O Ly$\alpha$/\ion{O}{vii} r           &$ 2.01 \pm 0.18 $ & $2.19 \pm 0.43 $&1.49 &1.97\\    
           Ne R-ratio                &$ 0.50 \pm 0.07 $ & $0.54 \pm 0.08 $&0.32 &0.31\\
    Ne G-ratio                       &$ 1.10 \pm 0.10 $ & $0.94 \pm 0.09 $&0.80 &0.75\\
     Ne Ly$\alpha$/\ion{Ne}{ix} r     &$ 0.27 \pm 0.04 $ & $0.62 \pm 0.06 $&0.26 &0.49\\  
\ion{Fe}{xvii} (17.09\AA+17.05\AA)/16.78\AA & n. a. & $3.32 \pm 0.88  $& 2.25& 2.22\\
           \ion{Fe}{xvii} 17.09\AA/17.05\AA & n. a. & $0.58 \pm 0.16  $& 0.81& 0.79\\   \hline
           red. $\chi^2$              &  $4.6$           & $1.4$     & &\\

\end{tabular}
\end{center}
\end{table*}
 
The available \emph{Chandra} data include a somewhat different wavelength interval. Only the triplets from O and Ne are covered, but the \emph{Chandra} MEG has better spectral resolution, allowing to reliably measure several iron lines. We use the three lines of \ion{Fe}{xvii} at 16.78~\AA{}, 17.05~\AA{} and 17.09~\AA{}, and calculate the ratio of the two doublet members and the total doublet to the third line (see Table~\ref{res_tab_chi_lines}). 
The best fit model for this data set has $v_0=575$~km~s$^{-1}$, slightly higher than found for the XMM-Newton data,  and $n_0=10^{12}$~cm$^{-3}$, with an unreduced value $\chi^2$ of 8.3 for 6 degrees of freedom. Again there is no difference between 6000~K and 10000~K taken as temperature for a black-body radiation background.
The neon G-ratio points again to lower infall velocities. Both fits have a tight lower bound on the density 
$n_0=10^{12}$~cm$^{-3}$ of the infalling gas. 
We estimate the error as one grid point, i.e.,  $\pm 25$~km~s$^{-1}$ for $v_0$ and $\pm0.33$ for $\log(n_0)$. The free-fall velocity for TW~Hya is between 500~km~s$^{-1}$ to 550~km~s$^{-1}$, setting a tight upper bound and we adopt $v_0=525$~km~s$^{-1}$ as the best value.  We thus conclude that  the same shock model is capable of explaining the line ratios observed in both the XMM-Newton RGS and {\it Chandra} HETGS grating spectra of TW~Hya.

\subsubsection{Global fit}

\label{globalfit}In our second step we implement the shock emission as XSPEC table model and proceed with a
normal XSPEC analysis of the medium and high resolution XMM-Newton spectra.
The analysis of both the {\it Chandra} and
XMM-Newton data \citep{2002ApJ...567..434K,twhya} suggested the presence of
higher temperature plasma possibly from an active corona; the flare observed
in the \emph{Chandra} observation may be a signature of activity.

To account for these additional contributions
we add up to three thermal VAPEC models and include interstellar extinction.
These additional thermal components are calculated in the low density limit and are meant to represent a coronal contribution. 
The elemental abundances for all components are coupled. It would be interesting to check if the accreting plasma is grain depleted compared to the corona, but the data quality is not sufficient to leave more abundances as free fit parameters. Because of the parameter degeneracy between the emission measure of the cool component and the interstellar absorption column $N_H$, we kept the latter fixed at $N_H=3.5\times10^{20}$~cm$^{-2}$, a value suggested by \citet[][ where a detailed discussion is given]{Rob0507}. The fit uses the data from one of the \emph{XMM} EPIC MOS detectors, RGS1 and RGS2 and \emph{Chandra's} HEG and MEG simultaneously in the energy range from 0.2~keV to 10~keV. The normalisation between the instruments is left independent to allow for calibration uncertainties and possible brightness variations between the observations. Because of the much larger count rates the lower resolution MOS detectors tend to dominate the $\chi^2$ statistic. To balance this we include all available grating information, but only one (MOS1) low resolution spectrum in the global fit.

\begin{figure}
\resizebox{\hsize}{!}{\includegraphics[angle=-90]{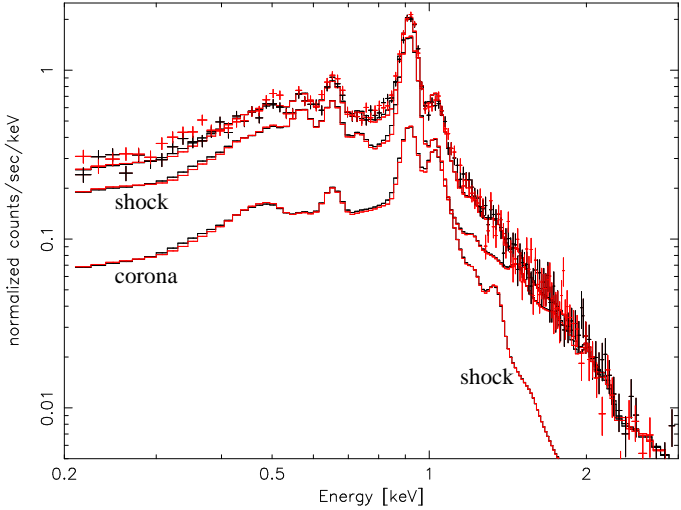}}
\caption{\label{epic}Data for the MOS1 (black symbols) and the MOS2 (red/grey symbols) together with our best fit model~C and its components}
\end{figure}

\begin{figure}
\resizebox{\hsize}{!}{\includegraphics[angle=-90]{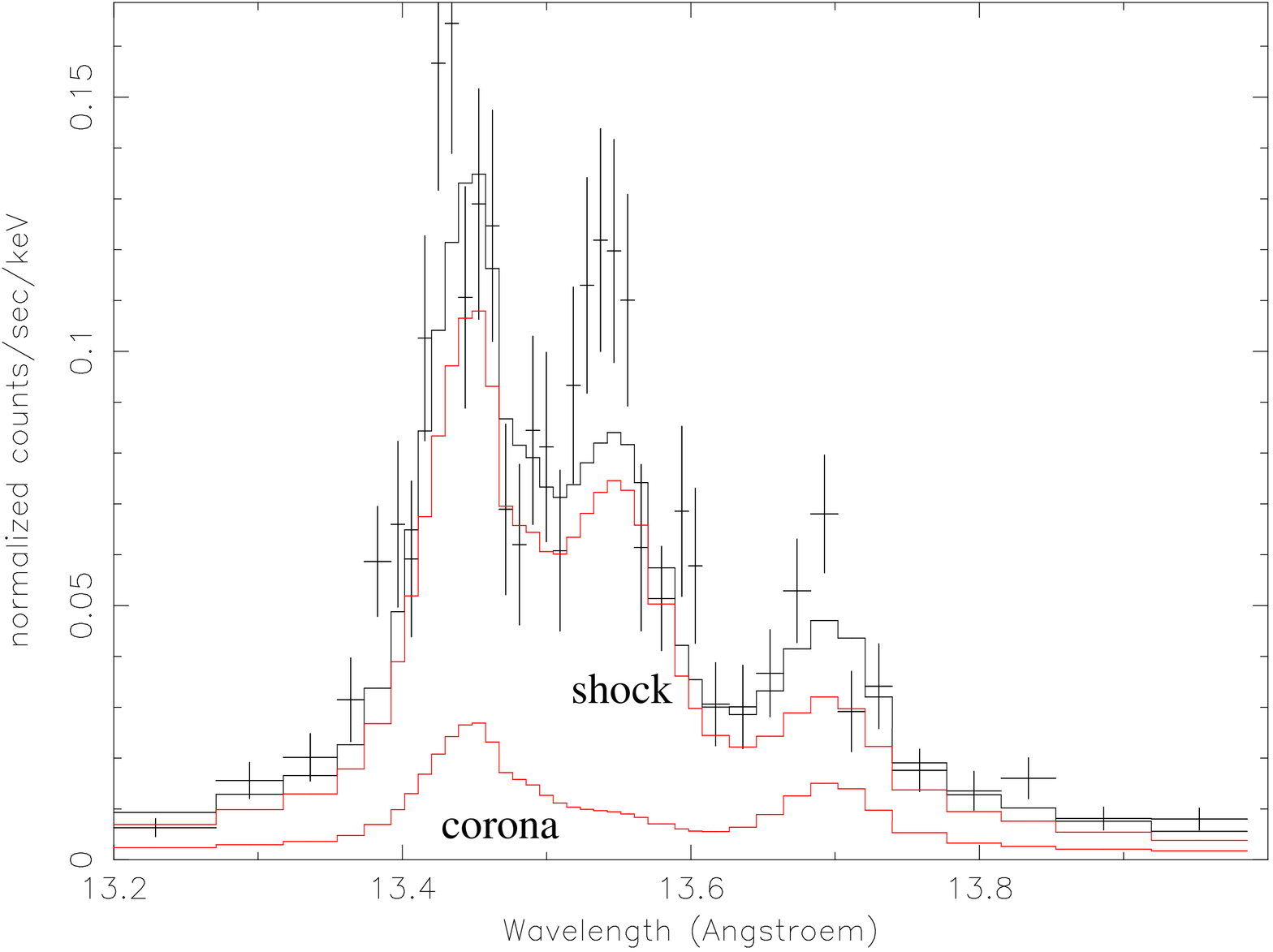}}
\caption{\label{netriplet}Ne triplet: Data from RGS2 with model~C (black) and accretion (red/grey upper line) and coronal component (red/grey lower line)}
\end{figure}

\begin{figure}
\resizebox{\hsize}{!}{\includegraphics[angle=-90]{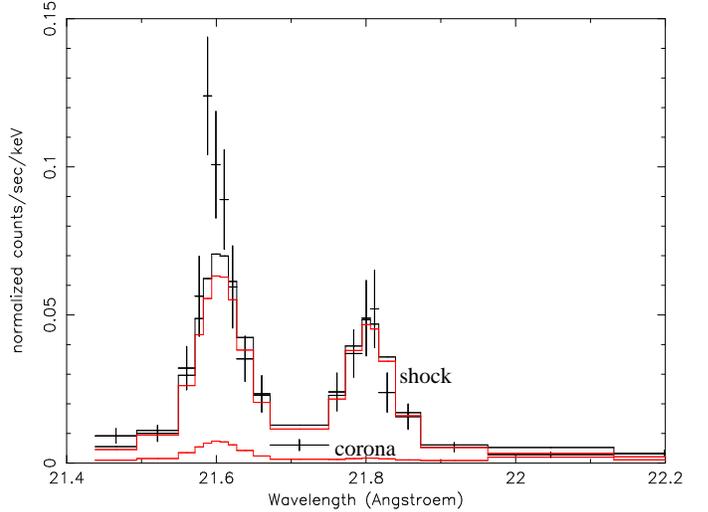}}
\caption{\label{otriplet}O triplet: Data from RGS1 with model~C (black) and accretion (red/grey upper line) and coronal component (red/grey lower line)}
\end{figure}

The fit results for our different models are presented in Table~\ref{chi}. 
Model A represents the best fitting pure accretion shock, while models B-C include additional
temperature components; the main improvement is obviously brought about by the
introduction of a high temperature ($kT \approx 1.30$~keV) component, representing 
the emission from a hot corona.  "Normal" coronae usually have emission measure distributions that can be described by a two-temperature model \citep{2003MNRAS.345..714B} and this motivates us to add cool low-density plasma component in model~C. Although the reduced $\chi^2$ is only marginally smaller, we regard this as a better model, because it can be naturally interpreted in terms of a stellar corona. A third low-density component does clearly not improve the fit any further (model D).  In
Fig.~\ref{epic} we show the recorded EPIC MOS1 low resolution spectrum, our best-fit model and separately the accretion and coronal contributions.  An inspection of Fig.~\ref{epic} shows that a energies below $\approx$~1.2 keV the overall emission is dominated
by the shock emission, while at higher energies the coronal contribution dominates
because of the thermal cut-off of the shock emission.  A 
distinction between high and low density plasma is possible only by examining the line ratios in the He-like triples or in iron lines. 
In broad band spectra cool coronal and shock plasma exhibit the same signatures. The ratio of their emission measures was therefore taken from the RGS modelling alone (model C and D).
Our global fit reproduces the triplet ratios quite satisfactorily as shown in Fig.~\ref{netriplet} for neon and in Fig.~\ref{otriplet} for the oxygen triplet.

The resulting infall velocity for model C is $530^{+8}_{-4}$~km~s$^{-1}$ and the density $n_0=(1\times10^{12}\pm 3\times10^{11})$~cm$^{-3}$ (errors are statistical only). In this scenario the total flux is dominated by the accretion shock ($3.7\times10^{-12}$~ergs~cm$^{-2}$~s$^{-1}$) which is about four times stronger than the cool corona ($1.0\times10^{-12}$~erg~cm$^{-2}$~s$^{-1}$) and five times stronger than the hot corona ($0.8\times10^{-12}$~erg~cm$^{-2}$~s$^{-1}$) in the 0.3-2.5~keV band.
\begin{table}
\caption{\label{chi}Reduced $\chi^2$ values for models with zero to three VAPEC components with temperature $kT_1$ to $kT_3$ in [keV] and one shock component. The absorption column is given in units of $[10^{20}\frac{1}{\mathrm{cm}^2}]$. The shock model converges to  $v_0 \approx 530$~km~s$^{-1}$ and $n_0 \approx 10^{12} \mathrm{ cm}^{-3}$ in all cases except A  ($v_0$ pegs at the free-fall velocity of $\approx 575$~km~s$^{-1}$). }
\begin{center}
\begin{tabular}{cccccc}
\hline \hline
Model  & $kT_1$ & $kT_2$& $kT_3$&$N_H$ & red. $\chi^2$ (dof)\\
\hline
A       & -     & -     & -     & =3.5  &2.8 (584) \\
B       & -     & -     & 1.33  & =3.5  &1.63 (580) \\
C       & 0.27  & -     & 1.35  & =3.5  &1.57 (577) \\
D       & 0.27  & 0.72  & 1.26  & =3.5  &1.56 (573) \\
\hline
\end{tabular}
\end{center}
\end{table}

\subsubsection{Elemental abundances}
\label{res_tw_abund}
The abundance fitting has to be performed recursively until the abundances converges, because different abundances lead to different cooling functions and thus change the whole shock structure.  Specifically,
we start from the set of elemental abundances determined by \citet{Rob0507} and iterate.
Our global fit procedure yields abundance values (Table~\ref{abund}) relative to solar abundances from \cite{1998SSRv...85..161G}, the errors given are purely statistical ($1\sigma$~range), while we believe the systematic error to be about 15\%. 
As a cross-check we compare the intensities of lines from different elements in our pure shock model from Sect.~\ref{res_tw_nv} and find that the abundance ratios estimated in this way roughly agree. 
The final abundance estimates (Table~\ref{abund}) show a metal depleted plasma with the exception of neon, which is enhanced by about a factor of ten compared to the other elements and nitrogen, which is enhanced by a factor of two. 
\begin{table}\caption{\label{abund}Abundance of elements relative to \protect{\citet{1998SSRv...85..161G}} and first ionisation potentials (FIP). The errors are statistical only, we estimate the systematic error to 15\%.}
\begin{center}
\begin{tabular}{ccc}
\hline\hline
Element & abundance & FIP [eV]\\
 \hline
C   & $0.20^{+0.03}_{-0.03}$ &11.3\\
N   & $0.51^{+0.05}_{-0.04}$ &14.6\\
O   & $0.25^{+0.01}_{-0.01}$ &13.6\\
Ne  & $2.46^{+0.06}_{-0.04}$ &21.6\\
Mg  & $0.37^{+0.10}_{-0.06}$ &7.6\\
Si  & $0.17^{+0.07}_{-0.07}$ &8.1\\
S   & $0.02^a$         &10.4\\
Fe  & $0.19^{+0.01}_{-0.01}$ &7.9\\
\hline
\end{tabular}
\end{center}
$^a$ formal $2\sigma$ limit
\end{table}
Metal depletion has also been observed in the wind of TW~Hya by \citet{2004AstL...30..413L} and was noted by \citet{twa5} using X-ray observations of the non-accreting quadruple system TWA~5 in the vicinity of TW~Hya. \citet{twhya} interpret the abundances as a sign of grain depletion, where the grain forming elements condensate and mainly those elements are accreted, which stay in the gas phase like the noble gas neon.  This is discussed in more detail by \citet{2005ApJ...627L.149D}, who collect evidence that metal depletion can be also seen in the the infrared and UV, where the spectral distribution indicates well advanced coagulation into larger orbiting bodies, which may resist the inward motion of the accreted gas. On the other hand, stars with active corona often show an enhancement of elements with a high first ionisation potential (IFIP), which also leads to an enhanced neon abundance \citep{Brinkman_HR_1099}.

\subsubsection{Filling factor and mass accretion rate}

\label{res_tw_fm} A comparison of the observed energy flux $f_{\mathrm{obs}}$ (at the distance $d$ to the star) and the simulated flux $f_{\mathrm{sim}}$ per unit area allows to calculate the accretion spot size $A_{\mathrm{spot}}$ through
\begin{equation} \label{res_aspot}
A_{\mathrm{spot}}=\frac{f_{\mathrm{obs}}(d)}{f_{\mathrm{sim}}(R_*)}4\pi d^2 \ .\end{equation}
The filling factor $f$ is the fraction of the stellar surface covered by the spot:
\begin{equation} \label{res_f}
f=\frac{A_{\mathrm{spot}}}{4\pi R_*^2}=\frac{f_{\mathrm{obs}}(d)}{f_{\mathrm{sim}}(R_*)}\frac{d^2}{R_*^2} \end{equation}
and the mass accretion rate is the product of the spot size and the mass flux per unit area, which in turn is the product of the gas density $\rho_0=\mu m_{\mathrm{H}} n_0$ and the infall velocity $v_0$:
\begin{equation} \label{res_mdot}
\frac{dM}{dt}=A_{\mathrm{spot}}\rho v_0=A_{\mathrm{spot}} \mu m_{\mathrm{H}} n_0 v_0 \ .\end{equation}
We assume that half of the emission is directed outward and can be observed, 
the other half is directed inwards, where it is absorbed.   For model C
the spot size is $1\times 10^{20}$~cm$^2$, yielding filling factors of 0.15\% and 0.3\% for $R_*=1\ R_{\sun}$ and $R_*=0.8\ R_{\sun}$ respectively and 
accretion rates of about $2\times10^{-10}\ M_{\sun}\; \mathrm{ yr}^{-1}$. 

\subsection{Shock position in the stellar atmosphere}

We now compare the ram pressure of the infalling gas to the stellar atmospheric pressures as calculated from \texttt{PHOENIX}; we specifically use
a density profile from AMES-cond-v2.6 with effective temperature 
$T_\mathrm{eff}=4000$~K, surface gravity $\log g=4.0$ and solar metalicity (\citet{2005tdug.conf..565B} based on \citet{2001ApJ...556..357A}). The chosen 
stellar parameters resemble those of typical CTTS. The shock front is expected to form, where the ram pressure approximately equals the stellar atmospheric pressure, which increases exponentially inwards. 
In a strict 1D-geometry photons emitted upwards out of  the cooling zone will be absorbed by an infinite accretion column, but in a more realistic geometry they can pass either through the stellar atmosphere or the pre-shock gas as can be seen in Fig.~\ref{funnel_names}. We estimate a lower limit for the hydrogen column density of $N_H=10^{20}$~cm$^{-2}$, the actually measured column density is  $3.5\times 10^{20}$~cm$^{-2}$ \citep{Rob0507} by adding the column density between shock front and emitting ion to the the column density of the pre-shock gas, which the photons penetrate before escaping from the accretion funnel outside the stellar atmosphere. The optical depth of the stellar atmosphere is far higher than our lower limit for the pre-shock gas. This estimate proves that shocks, as described by our model, are actually visible; for a contrary view on this subject matter see \citet{2005CSSS.519.D}.

\section{Discussion}
\label{dis}
\begin{figure}
\resizebox{\hsize}{!}{\includegraphics{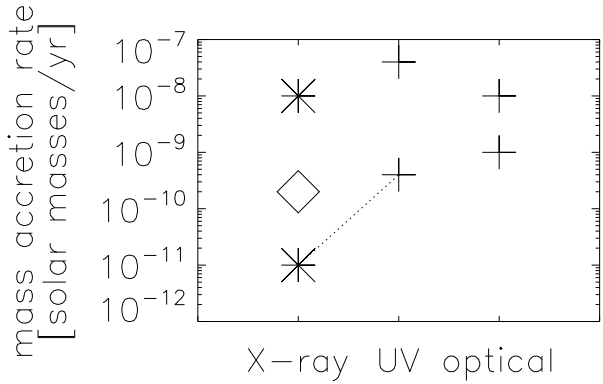}}
\caption{\label{dis_mdot_fig}Estimated mass accretion rates sorted by the energy of the observation. +: optical and UV measurements; *: X-ray, for the data from \protect{\citet{twhya}} the dotted line connects to the UV measurement, which delivers the filling factor, $\diamondsuit$: this simulation; data sources: see text}
\end{figure}

The best fit parameters of our shock model to match the X-ray observations of TW~Hya are obtained by using a shock with the parameters $v_0\approx525$~km~s$^{-1}$ and $\log n_0\approx12$. Previously infall velocities closer to $\sim300$~km~s$^{-1}$ were reported by a number of authors \citep{lamzin,calvetgullbring}.  Other observational evidence also suggests lower values; 
in the UV emission is found in highly ionised emission lines (\ion{C}{iv}, \ion{N}{v} and \ion{O}{vi}) extending up to $\approx 400$~km~s$^{-1}$ and in cool ions in absorption (\ion{Fe}{ii}) against a hot continuum likely emerging from an accretion  spot \citep{2004AstL...30..413L}. 
Since the gas strongly accelerates close to the stellar surface, the density will be lower in the high velocity region because of particle number conservation, so, depending on the geometry of the accretion funnel, the emission measure in this region may well be very small. In this case the observed lines will have weak wings extending to larger velocities, which are difficult to identify observationally. We therefore regard these observations only as a lower bound; an upper bound is provided by the free-fall velocity of $\sim500-550$~km~s$^{-1}$. \\
Measurements of TW~Hya in different wavelength regions lead to conspicuously distinct mass accretion rates. Generally, the published estimates far exceed the results of our simulation, which gives an accretion rate of $\approx (2\pm0.5)\times10^{-10}\ M_{\sun}\; \mathrm{ yr}^{-1}$ and filling factors of 0.2\%-0.4\%. \citet{2002ApJ...571..378A} and \citet{2002ApJ...580..343B} use optical spectroscopy and photometry and state a mass accretion rate between $10^{-9}$ and $10^{-8}\ M_{\sun} \;\mathrm{ yr}^{-1}$ and a filling factor of a few percent. In the UV the picture is inconsistent. On the one hand, two empirical relations for line intensities as accretion tracers \citep{2000ApJ...539..815J} indicate mass accretion rates above $3\times 10^{-8}\ M_{\sun}\; \mathrm{ yr}^{-1}$ (data from \citet{2000ApJS..129..399V}, evaluated by \citet{2002ApJ...567..434K}), on the other hand fitting blackbodies on the UV-veiling by \citet{2000ApJ...535L..47M} suggests a significantly lower value: $4\times 10^{-10}\ M_{\sun}\; \mathrm{ yr}^{-1}$. 
A similar procedure has been earlier applied by \citet{2000A&A...354..621C} with a much larger filling factor. The previous X-ray analyses by \citet[][ $\dot{M}=10^{-8}\ M_{\sun}\; \mathrm{ yr}^{-1}$]{2002ApJ...567..434K} and \citet[][ $\dot{M}=10^{-11}\ M_{\sun}\; \mathrm{ yr}^{-1}$]{twhya} suffer from the problem that they use filling factors extracted from UV-measurements (from \citet{2000A&A...354..621C} and  \citet{2000ApJ...535L..47M} respectively) and a post-shock density calculated from X-ray observations which does not necessarily represent the same region. Our simulation now is the first attempt to rely solely on the X-ray measurements. The different values for mass accretion rates are summarised in Fig.~\ref{dis_mdot_fig}.\\
A physical explanation for these differences goes as follows: At longer wavelengths one observes plasma at cooler temperatures and in general the filling factor and mass accretion rates are higher, because the spot is inhomogeneous with different infall velocities. Fast particles would be responsible for a shock region with high temperatures which is observed in X-rays, whereas in other spectral bands cooler areas can be detected, and therefore the total area and the observed total mass flux is larger. Accretion spots with these properties are predicted by the magneto-hydrodynamic simulations recently performed by \citet{2004ApJ...610..920R} and \citet{2006MNRAS.371..999G}. 
Very probably some of the difference can be attributed to intrinsically changing accretion rates. Simulations of the inner flow region often show highly unstable configurations \citep{2006AN....327...53V}.

We showed that basic properties of the X-ray spectra from CTTS can be naturally explained by accretion on a hot spot, but this is not the only X-ray emission mechanism. 
To understand especially the high energy tail we had to introduce two thermal components which fit a corona as it is expected in late-type stars and suggested by observations of activity; in the case of TW~Hya the shock can dominate the overall X-ray
emission. 
Forthcoming high-resolution observations will hopefully allow to extend the sample of CTTS, where a similar analysis is feasible.
\begin{acknowledgements}
CHIANTI is a collaborative project involving the NRL (USA), RAL (UK), MSSL (UK), the Universities of Florence (Italy) and Cambridge (UK), and George Mason University (USA).\\
H.M.G., J.R and C.L. acknowledge support from DLR under 50OR0105.
\end{acknowledgements}
\bibliographystyle{aa} 
\bibliography{../articles}
\end{document}